\documentclass{article}
\usepackage{spconf,amsmath,graphicx}
\usepackage{booktabs} 

\usepackage{makecell}
\usepackage{subfigure}

\usepackage{enumitem}
\setlist{nosep, leftmargin=14pt}

\usepackage{url}
\usepackage[table,dvipsnames]{xcolor}
\usepackage{amssymb}
\definecolor{LightGreen}{HTML}{a2ffa6}
\definecolor{LightRed}{HTML}{ff9191}

\newcommand{\pbad}{\cellcolor{LightRed}}
\newcommand{\pgood}{\cellcolor{LightGreen}}

\title{Parameter choices in HaarPSI for IQA with medical images}
\name{
\begin{tabular}{@{}c@{}}
Clemens Karner$^{1}$ \quad Janek Gröhl$^{2}$ \quad Ian Selby$^{3,4}$ \quad Judith Babar $^{3,4}$ \quad Jake Beckford$^{4}$ \quad Thomas R Else$^{2}$ \\ Timothy J Sadler$^{4}$ \ \  Shahab Shahipasand$^{4}$ \ \ Arthikkaa Thavakumar$^{4}$ \ \ Michael Roberts$^{5}$ \\ James H.F. Rudd$^{6}$ \quad Carola-Bibiane Schönlieb$^{5}$ \quad Jonathan R Weir-McCall$^{7}$ \quad Anna Breger$^{1,5,*}$
\end{tabular}
}
\address{\normalsize{$^{1}$ Medical University of Vienna, Center for Medical Physics and Biomedical Engineering, Vienna, Austria }\\
     \normalsize{$^{2}$ University of Cambridge, Department of Physics, Cambridge, UK}\\
     \normalsize{$^{3}$ University of Cambridge, Department of Radiology, Cambridge, UK}\\
     \normalsize{$^{4}$ Cambridge University Hospitals, Department of Radiology, Cambridge, UK}\\
          \normalsize{$^{5}$ University of Cambridge, Department of Applied Mathematics and Theoretical Physics, Cambridge, UK}\\
     \normalsize{$^{6}$ University of Cambridge, Heart \& Lung Research Institute, Cambridge, UK}\\
     \normalsize{$^{7}$ King's College London, School of Biomedical Engineering \& Imaging Sciences, London, UK}\\
     \normalsize{$^\star$Corresponding author. E-mail: ab2864@cam.ac.uk}}
     
\begin{document}
\maketitle
\begin{abstract}
When developing machine learning models, image quality assessment (IQA) measures are a crucial component for the evaluation of obtained output images. However, commonly used full-reference IQA (FR-IQA) measures have been primarily developed and optimized for natural images. In many specialized settings, such as medical images, this poses an often overlooked problem regarding suitability. In previous studies, the FR-IQA measure HaarPSI showed promising behavior regarding generalizability. The measure is based on Haar wavelet representations and the framework allows optimization of two parameters. So far, these parameters have been aligned for natural images. Here, we optimize these parameters for two medical image data sets, a photoacoustic and a chest X-ray data set, with IQA expert ratings. We observe that they lead to similar parameter values, different to the natural image data, and are more sensitive to parameter changes. We denote the novel optimized setting as HaarPSI$_{MED}$, which improves the performance of the employed medical images significantly ($p<0.05$). Additionally, we include an independent CT test data set that illustrates the generalizability of HaarPSI$_{MED}$, as well as visual examples that qualitatively demonstrate the improvement. The results suggest that adapting common IQA measures within their frameworks for medical images can provide a valuable, generalizable addition to employment of more specific task-based measures.
\end{abstract}
\section{Introduction}
\label{sec:intro}
In the last decade, tremendous progress has been made in the development of advanced machine learning models for various tasks. The evaluation of models employing large image data sets needs automation, where image quality assessment (IQA) measures are suitable candidates. IQA measures quantitatively compute the quality of an image and can roughly be divided into no-reference IQA (NR-IQA), which scores the quality of an image on its own, and full-reference IQA (FR-IQA), which quantifies the similarity between two images through a notion of distance. 

Two of the most commonly used FR-IQA measures, peak signal-to-noise ratio (PSNR)~\cite{639240} and structural similarity index measure (SSIM)~\cite{live2_ssim}, have been known for more than 20 years. However, PSNR and SSIM underperform on several specialized tasks, including medical imaging~\cite{breger2024study1}. Today, there exists a wide range of FR-IQA measures, including measures based on SSIM~\cite{iwssim, msssim} and recent advancements based on machine learning, such as the learned perceptual image patch similarity (LPIPS)~\cite{lpips} and the deep image structure and texture similarity (DISTS)~\cite{9298952}. When selecting an IQA measure, it is important to consider the type of image evaluated as well as the intended application. Medical images often have very distinct properties depending on the modality and subsequent task. In contrast to natural images, which may focus on overall aesthetics, fine details and local perturbations are usually of great importance. As most FR-IQA measures have been developed and optimized primarily for natural images, it is not surprising, that there is a significant performance gap when applied to medical images~\cite{breger2024study2, IQAdylovmri}. 

In this paper, we study the generalizability of the Haar wavelet-based perceptual similarity index (HaarPSI)~\cite{Reisenhofer18} to medical images as it has shown promising results across domains \cite{IQAdylovmri, breger2024study2}.  HaarPSI is based on comparing the frequency decompositions of two images using Haar wavelets, as presented in Section \ref{methdata}. The framework includes two adjustable parameters, which have originally been optimized for the natural image data sets LIVE~\cite{live1,live2_ssim,live3}, TID2008~\cite{tid2008}, TID2013~\cite{tid2013} and CSIQ \cite{csiq}. Here, we build upon \cite{breger2024study2}, where the performance of a variety of FR-IQA and NR-IQA measures is tested on $4$ data sets, in particular, $2$ natural image data sets as well as a photoacoustic (PA)~\cite{Janek7,photoacoustic_dataset}
and a chest X-ray (CXR)~\cite{breger2024study2} data set with expert quality ratings. We study the impact of optimizing the HaarPSI parameters on the two medical data sets PA and CXR, comparing it with the default parameter settings on two natural image data sets and, furthermore, test it on an independent publicly available medical computed tomography (CT) data set~\cite{MICCAI2023}. 
We also include a comparison of the results to other commonly used FR-IQA measures, namely PSNR, SSIM, MS-SSIM~\cite{msssim}, IW-SSIM~\cite{iwssim}, GMSD~\cite{gmsd}, FSIM~\cite{fsim}, MDSI~\cite{mdsi}, LPIPS and DISTS. For visualization, we provide a qualitative comparison of HaarPSI, HaarPSI$_{MED}$, PSNR and SSIM on images of the CXR data set in Figure \ref{fig:CXR}, as well as an independent MRI example with synthetic degradations in Figure \ref{fig:toyexp}.  

Finally, we provide additional experiments and complementary information in the supplemental material (S.M.).  
\section{Methods and Data}\label{methdata}
HaarPSI is a FR-IQA measure based on comparing the Haar wavelet frequency decomposition of two images introduced in 2016, cf.~\cite{Reisenhofer18}. It is a simplification of the feature similarity index (FSIM), resulting in lower computational effort. HaarPSI computes quality values in the range $[0,1]$, with $1$ corresponding to the best value. It employs the one-dimensional low-pass scaling Haar filter $h_1^{1D}=\frac{1}{\sqrt{2}} \cdot[1,1]$ and a high-pass Haar filter $g_1^{1D}=\frac{1}{\sqrt{2}} \cdot [-1,1]$ as well as the one-dimensional filters $g_j^{1D}=h_1^{1D} \star (g_{j-1}^{1D})_{\text{\textuparrow}2}$ and $h_j^{1D}=h_1^{1D} \star (h_{j-1}^{1D})_{\text{\textuparrow}2}
$, where $\star$ denotes the convolution and $\text{\textuparrow}2$ the dyadic upsampling operator. These are used to construct the two-dimensional Haar filter for any scale $j \in \mathbb{N}$ as $
g_j^{(1)}=g_j^{1D} \otimes h_j^{1D}$ and $
g_j^{(2)}=h_j^{1D} \otimes g_j^{1D}
$.
For two grayscale images $f_1, f_2 \in l^2(\mathbb{Z}^2)$ HaarPSI is computed by
$$
\mathrm{HaarPSI}_{f_1,f_2}=l_{\alpha}^{-1}\left(\frac{\sum_{x}\sum_{k=1}^{2}\mathrm{HS}_{f_1,f_2}^{(k)}(x) \cdot \mathrm{W}_{f_1,f_2}^{(k)}(x)}{\sum_{x}\sum_{k=1}^{2}\mathrm{W}_{f_1,f_2}^{(k)}(x)}\right)^2,
$$
where $x$ represents the image pixels and for $k\in \{1,2\}$ we have $W^{(k)}_{f_1,f_2} := \max (W^{(k)}_{f_1},W^{(k)}_{f_2})$ as well as $W_f^{(k)}(x)=|(g_3^{(k)}\star f)(x)|$. 
$\mathrm{HS}_{f_1, f_2}^{(k)}$ is the local similarity map defined as
$$
\mathrm{HS}_{f_1, f_2}^{(k)}(x)=l_{\alpha} \bigl(\frac{1}{2}\sum_{j=1}^{2}S(|(g_{j}^{(k)}\star f_1)(x)|,|(g_{j}^{(k)}\star f_2)(x)|,C)\bigr).
$$
The HaarPSI framework allows the choice of the parameters $C > 0$ and $\alpha > 0$, which are part of the function 
\begin{equation}\label{eq:C}
\mathrm{S}(a,b,C)=\frac{2ab+C}{a^2+b^2+C},
\end{equation}
where $a,b>0$ and of the logistic function
\begin{equation}\label{eq:alpha}
l_{\alpha}(y)=\frac{1}{1+e^{-\alpha y}}, 
\end{equation}
with $y \in \mathbb{R}$.

For the parameter optimization, we employ a publicly available photoacoustic data set~\cite{photoacoustic_dataset} (PA, $1134$ grayscale images, $288 \times 288$ pixels, $2$ expert ratings) and a chest X-Ray data set~\cite{breger2024study2} (CXR, $1571$ grayscale images, up to $3512 \times 3756$ pixels, $5$ expert ratings). 
The impact of the parameter choices are compared on the publicly available natural image data sets \textit{Laboratory for Image \& Video Engineering Image Quality Assessment Database Release 2}~\cite{live3} (LIVE, $982$ color images, up to $768 \times 512$ pixels, ratings by on average $22$ people) and \textit{Laboratory for Image \& Video Engineering Multiply Distorted Image Quality Database}~\cite{livemulti2} (LIVE$_{M}$, $405$ color images, up to $1280 \times 720$ pixels, ratings by on average $18$ people). Moreover, for independent testing, we include the publicly available \textit{Low-dose Computed Tomography Perceptual Image Quality Assessment Grand Challenge Dataset} (CT, $272$ test grayscale images, $512 \times 512$ pixels, $6$ expert ratings), the reference images were kindly provided by the authors. We cropped the images to the region of interest (ROI, including the full anatomical structure, $470 \times 390$ pixels) to reduce the amount of background pixels, see S.M. section 2.

To assess the alignment of HaarPSI`s evaluation and manual expert ratings, we employ the Spearman Rank Correlation Coefficient (SRCC) and Kendall Rank Correlation Coefficient (KRCC), stating how strongly the ranks of the quality ratings by the graders and the computed IQA values correlate. First, we optimize the parameters $C$ and $\alpha$ (Eq.\ref{eq:C} and Eq.\ref{eq:alpha}) for the CXR and PA data individually. Following the original HaarPSI paper \cite{Reisenhofer18}, we use grid search with a precision of 4 digits and use the suggested ranges, i.e.~$C$ in $\{5, 6, \ldots, 100\}$ and $\alpha$ in $\{2,2.1, \ldots, 8\}$. The chosen parameters maximize the mean SRCC to the z-scored annotations for the PA and CXR data sets individually. Subsequently, we optimize the mean SRCC over the PA and CXR data sets and denote the identified parameter choice by HaarPSI$_{MED}$. The results of this novel setting are compared with the default HaarPSI and other commonly used FR-IQA measures on the PA, CXR, LIVE, LIVE$_{M}$ and the independent medical CT data set, where we additionally report the statistical significance of the difference between HaarPSI$_{MED}$ and the other tested measures. We compute the statistical significance using the \texttt{paired.r} function from the R library \texttt{psych} and for the IQA measures we employ implementations provided by their authors. 
Our HaarPSI PyTorch implementation is based on the original TensorFlow-based code and is made available on Github\footnote{\url{https://github.com/ideal-iqa/haarpsi-pytorch}}. It allows explicit parameter choices as well as providing default suggestions including HaarPSI$_{MED}$. 
\vspace{-0.2cm}
\section{Results}
\begin{figure}[h!]
\begin{minipage}[b]{1\linewidth}
  \centering
  \centerline{\includegraphics[width=4cm]{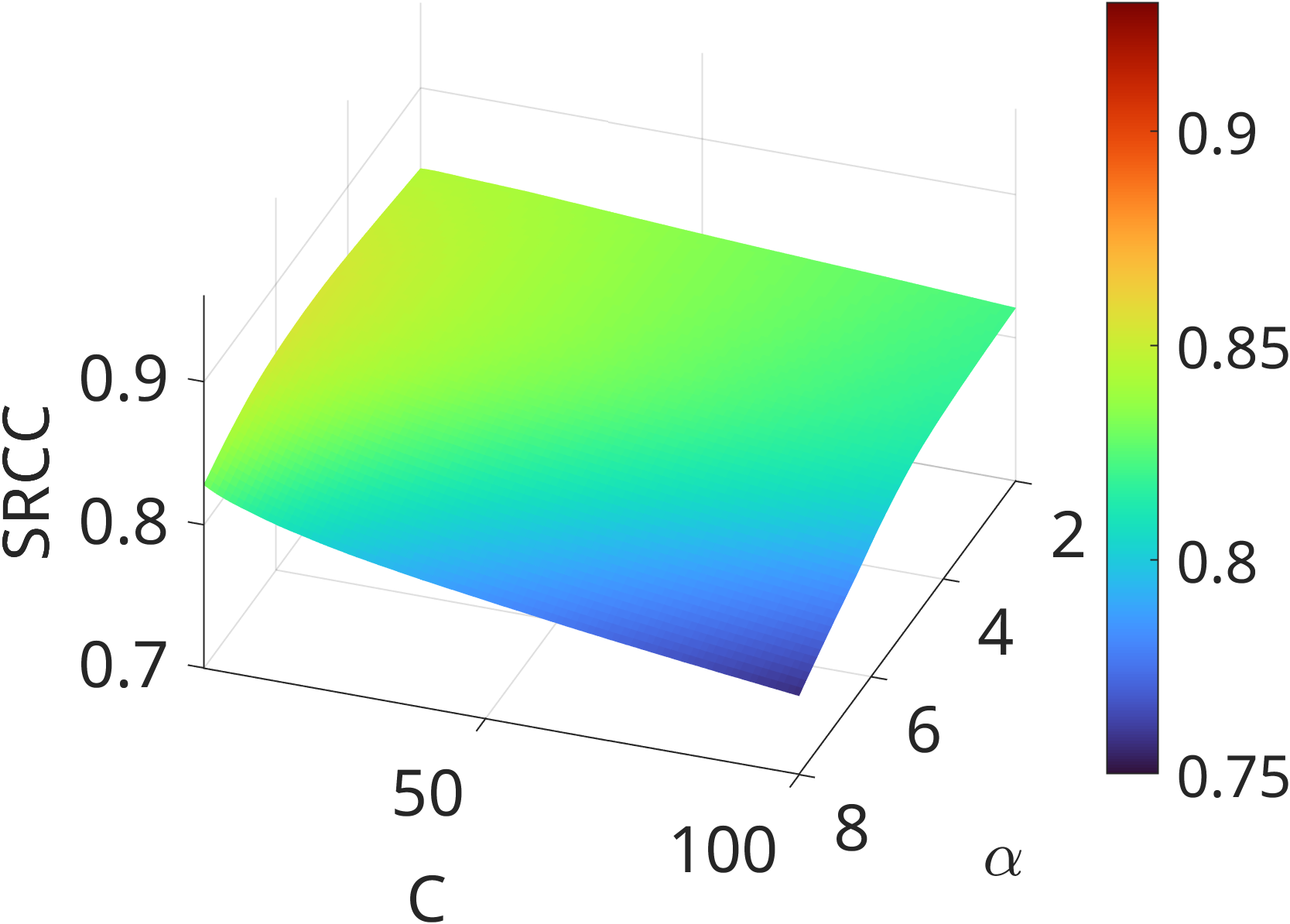}}
  \centerline{\small(a) Mean SRCC of CXR and PA}\medskip
\end{minipage}
\begin{minipage}[b]{.48\linewidth}
  \centering
\centerline{\includegraphics[width=3.5cm]{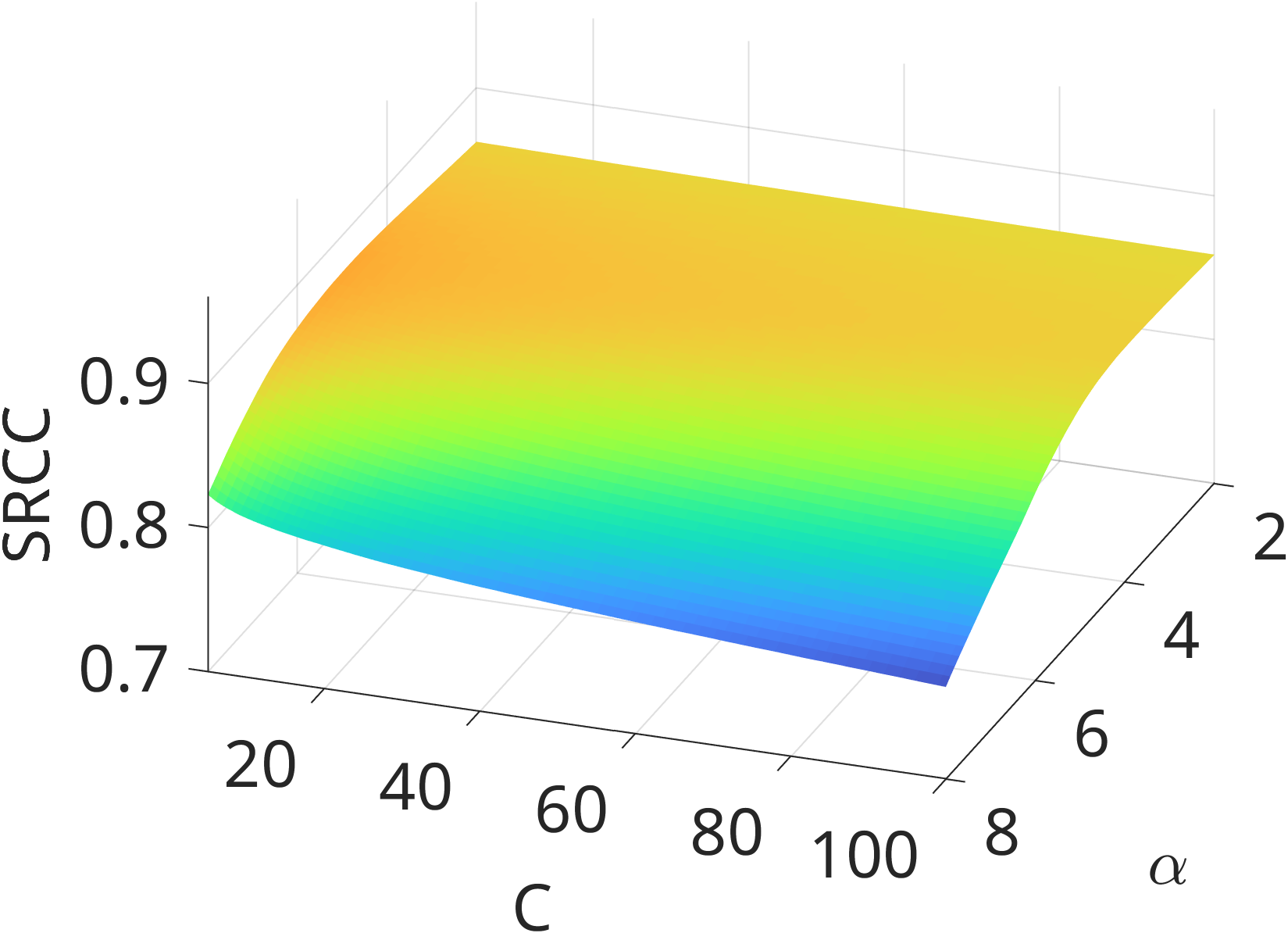}}
  \centerline{\small(b) CXR}\medskip
\end{minipage}
\hfill
\begin{minipage}[b]{0.48\linewidth}
  \centering
  \centerline{\includegraphics[width=3.5cm]{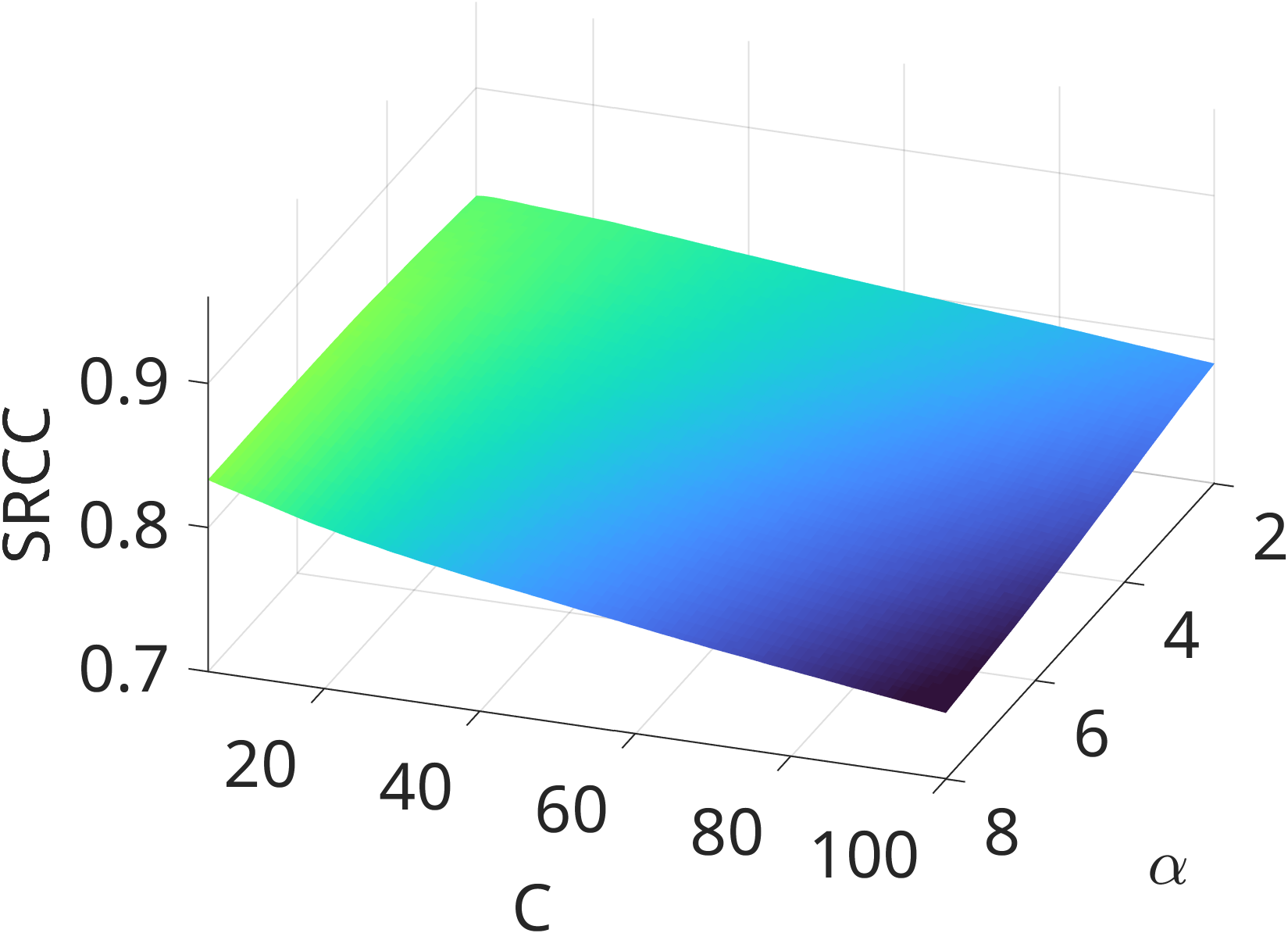}}
  \centerline{\small(c) PA}\medskip
\end{minipage}
\begin{minipage}[b]{0.48\linewidth}
  \centering
  \centerline{\includegraphics[width=3.5cm]{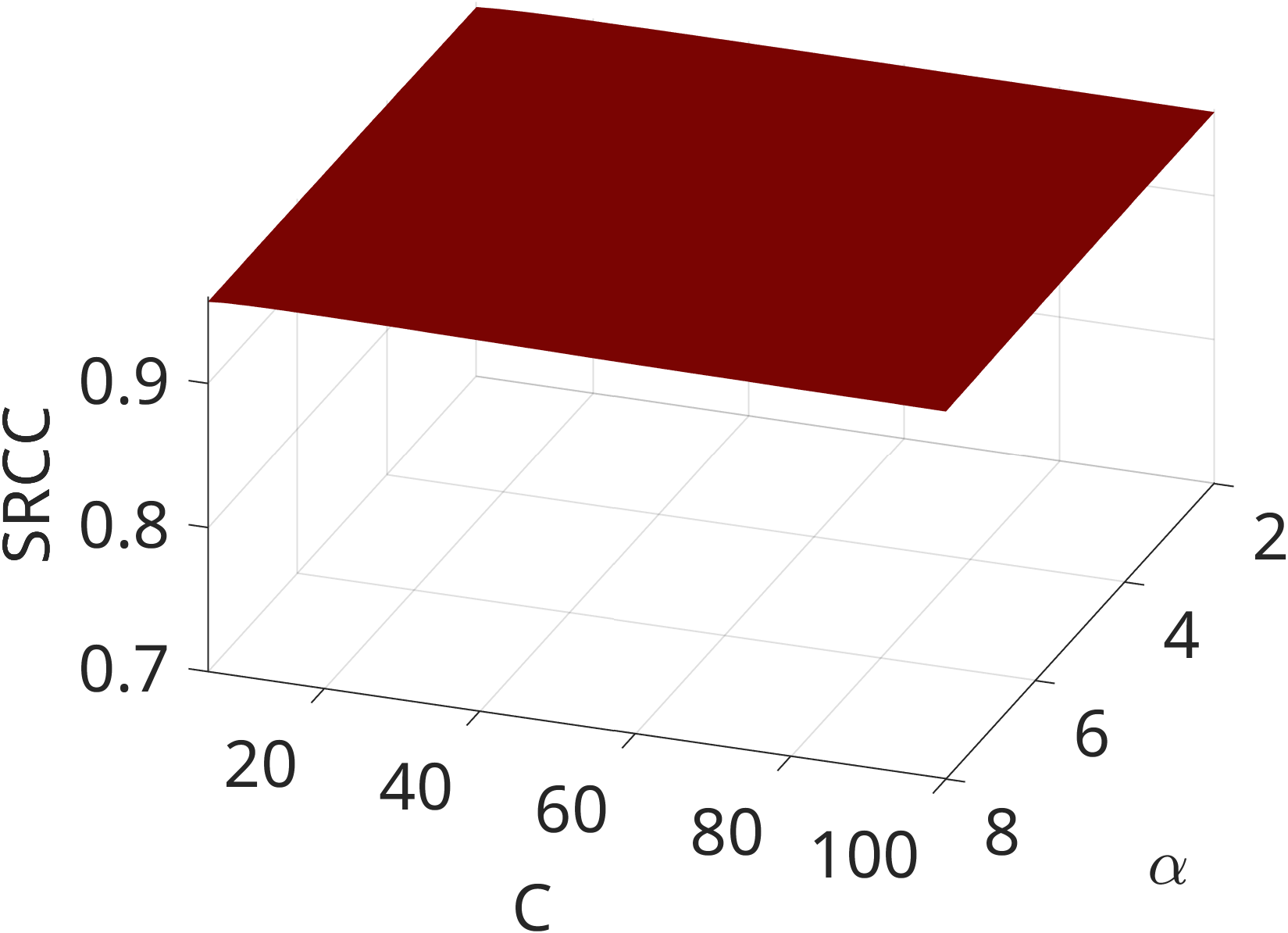}}
  \centerline{\small(f) LIVE}\medskip
\end{minipage}
\hfill
\begin{minipage}[b]{0.48\linewidth}
  \centering
  \centerline{\includegraphics[width=3.5cm]{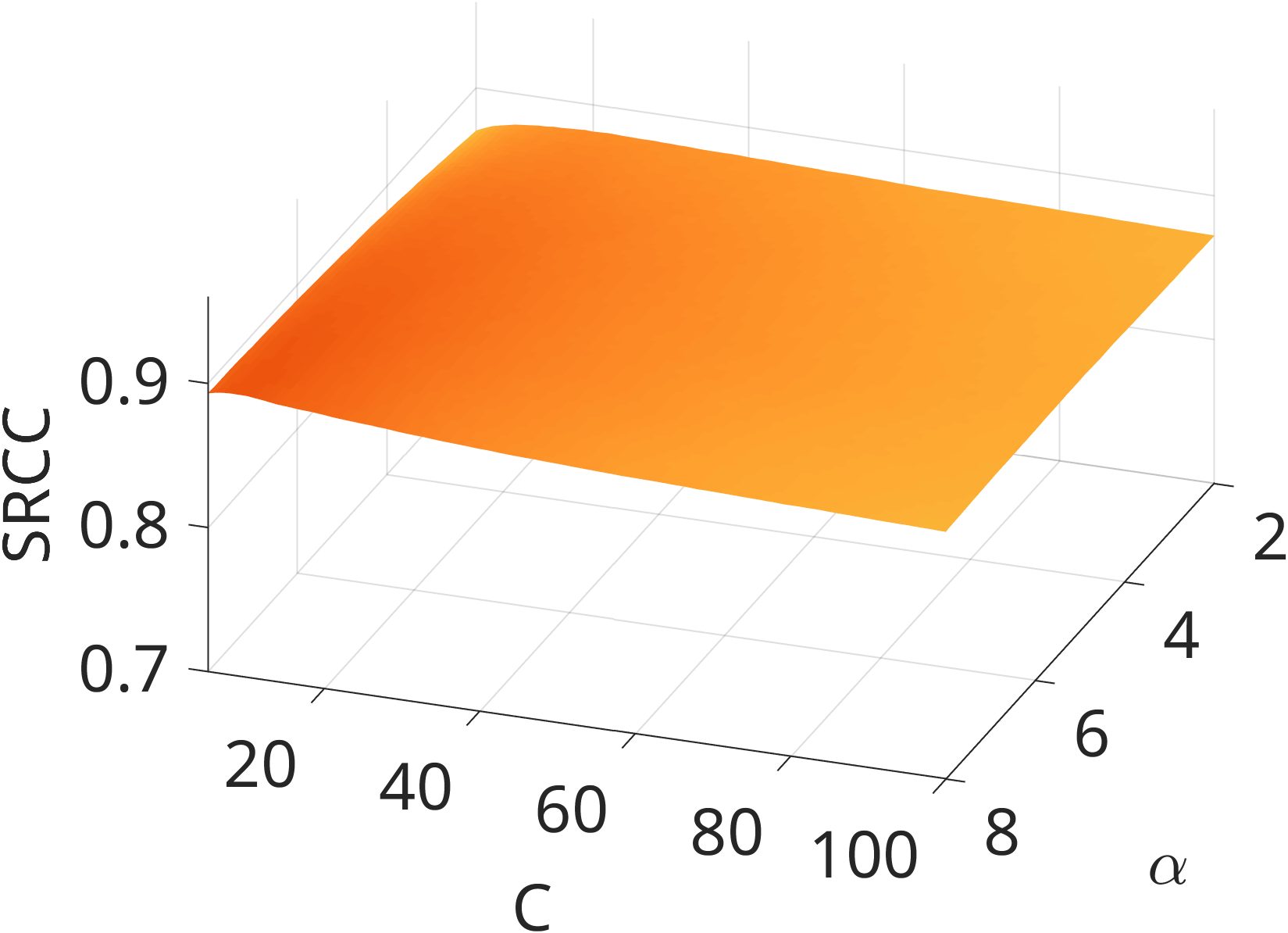}}
  \centerline{\small(g) LIVE$_{M}$}\medskip
\end{minipage}
\caption{\small SRCC values between the HaarPSI evaluation and manual quality ratings for the HaarPSI parameters $C$ in $\{5, 6, \ldots, 100\}$ and $\alpha$ in $\{2,2.1, \ldots, 8\}$.}
\label{fig:optimization}
\end{figure}

\begin{figure}
  \centering
  \subfigure[Reference]{ \includegraphics[width=0.14\textwidth]{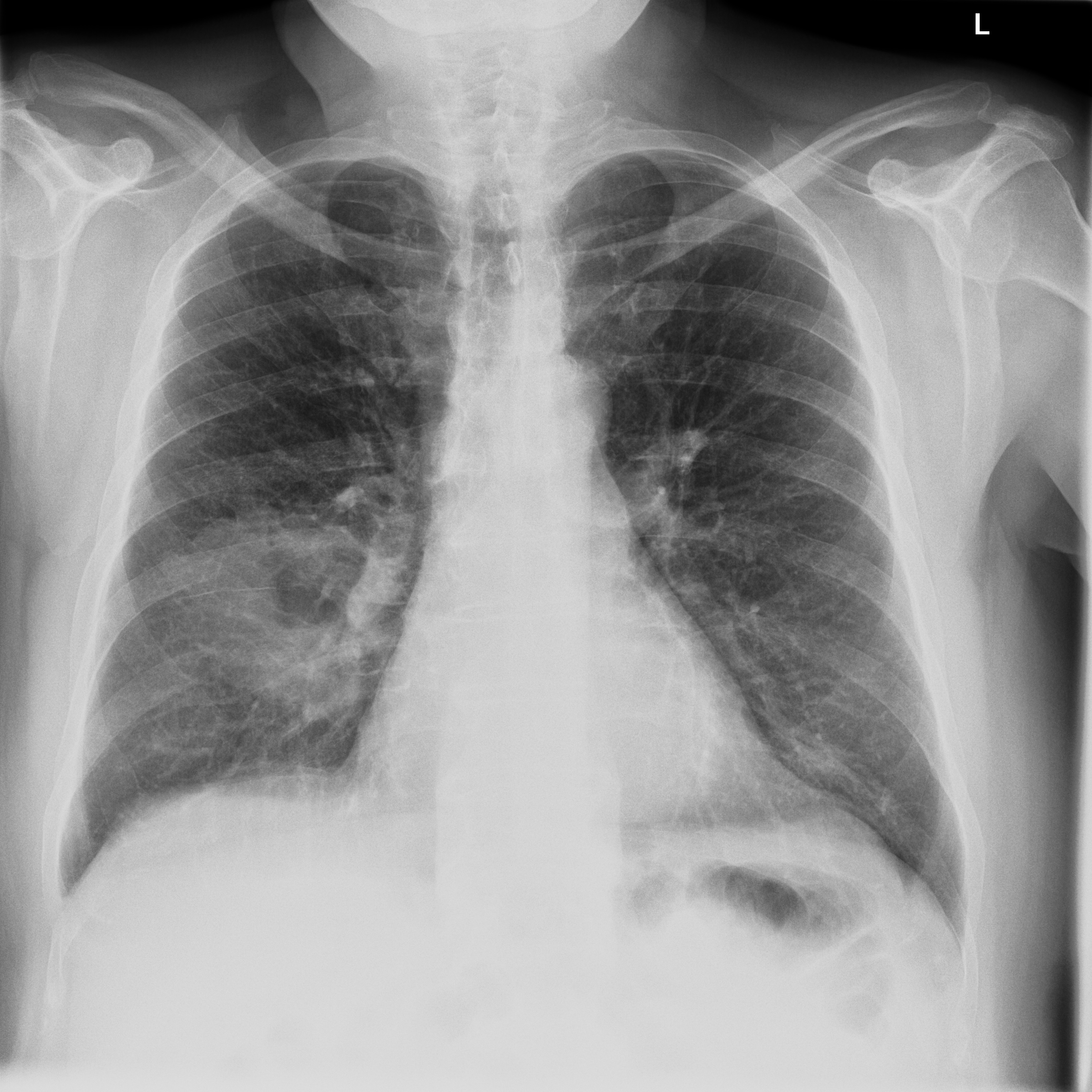} }
  \hfill
\subfigure[21.1/.90/.59/.56]{  \includegraphics[width=0.14\textwidth]{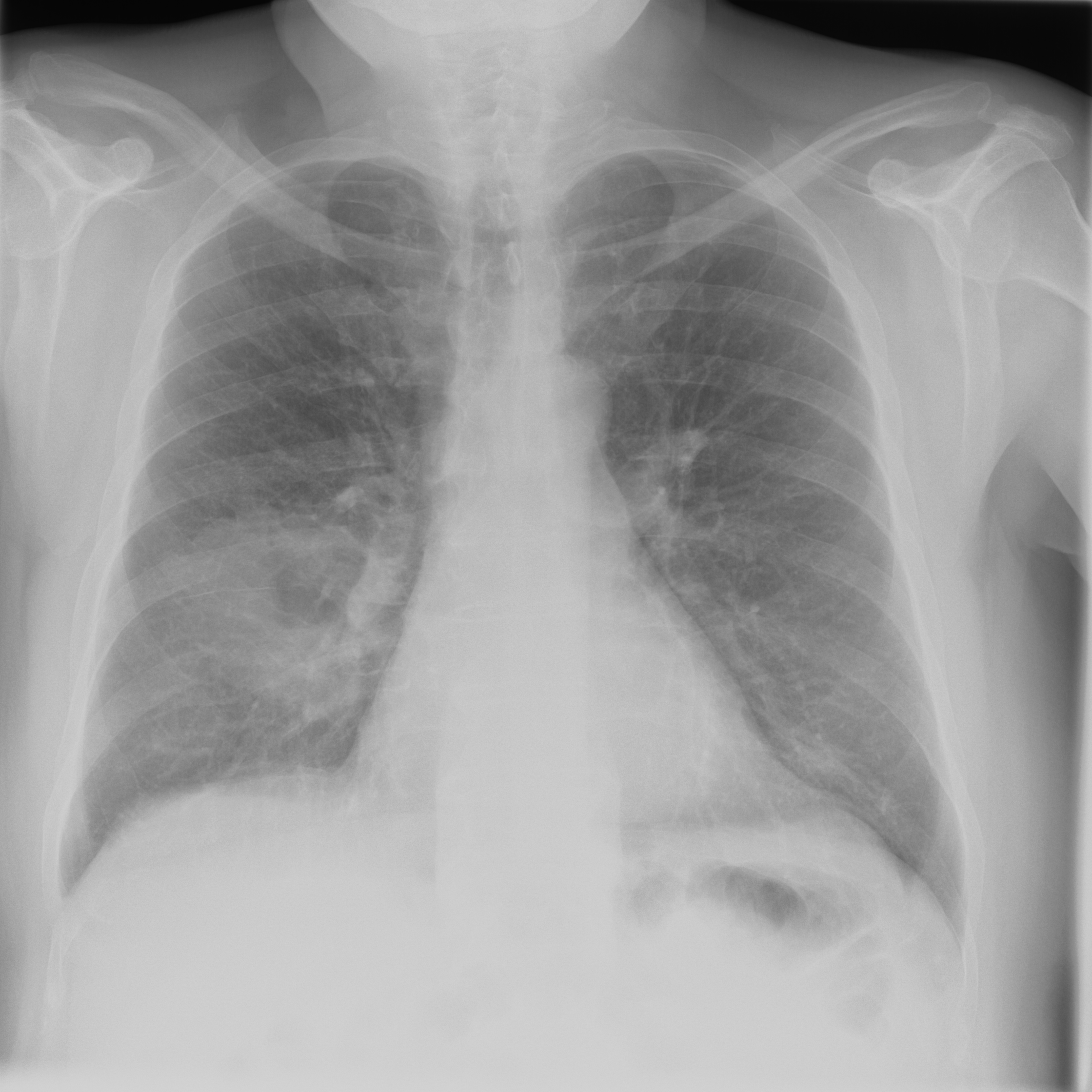}}
\hfill
 \subfigure[19.5/.88/.61/.59]{ \includegraphics[width=0.14\textwidth]{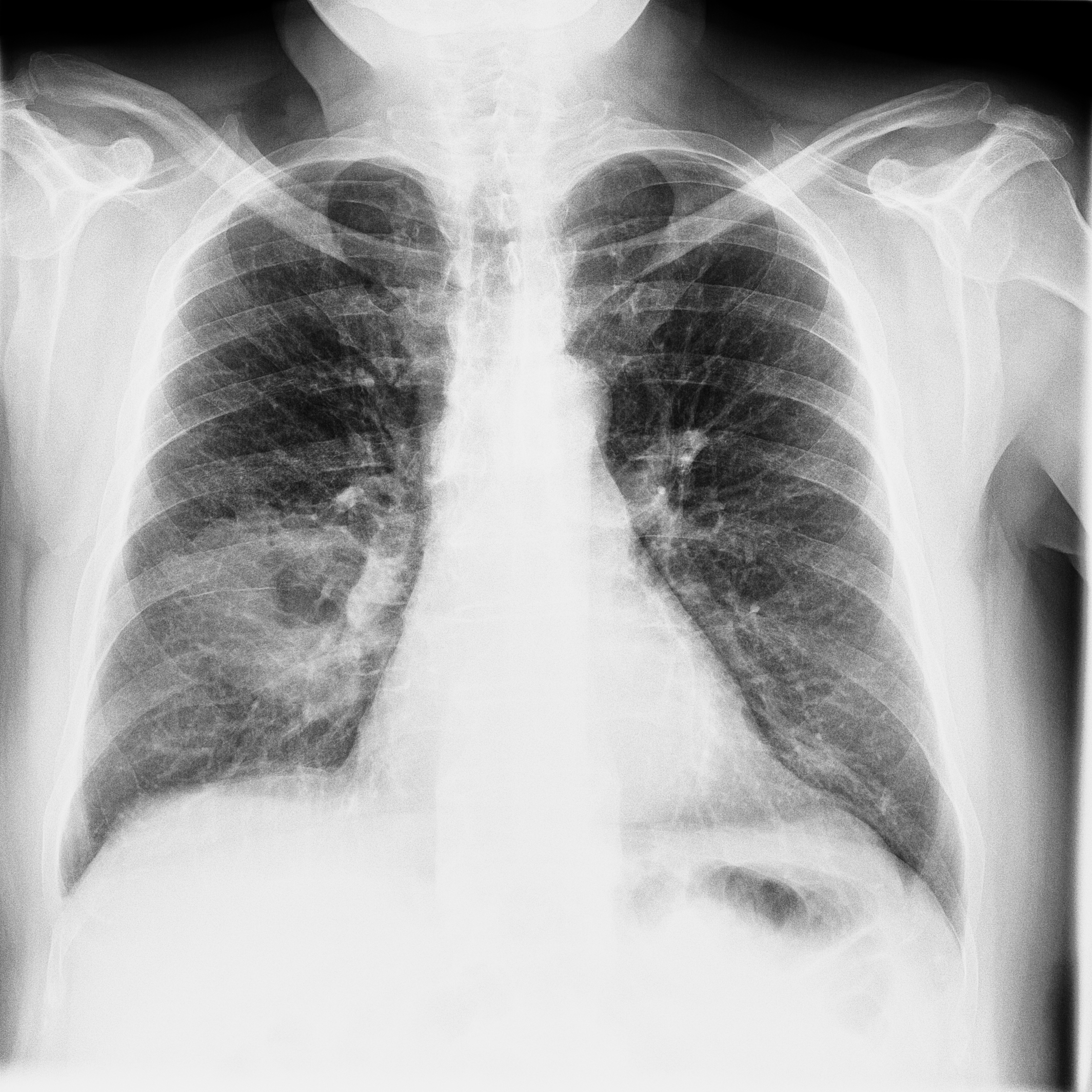}}
  \caption{\small CXR scans with different kinds of post-processing scored by PSNR/SSIM/HaarPSI/HaarPSI$_{MED}$. PSNR and SSIM  wrongly judge (b) to be the better image, whilst HaarPSI and HaarPSI$_{MED}$ correctly both score (c) as the better image. HaarPSI$_{MED}$ correctly identifies a bigger gap in quality.}
  \label{fig:CXR}
\end{figure}

\begin{figure}[t]
\centering 
\begin{minipage}[b]{0.3\linewidth}
  \centering
  \centerline{\includegraphics[width=2cm]{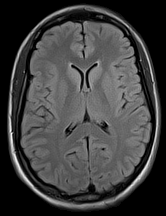}}
  \centerline{\footnotesize (a) Reference}\medskip
\end{minipage}
\hfill
\begin{minipage}[b]{0.3\linewidth}
  \centering
  \centerline{\includegraphics[width=2cm]{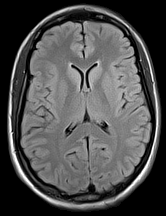}}
  \centerline{\footnotesize (b) 22.6/.97/.97/.96}\medskip
\end{minipage}
\hfill
\begin{minipage}[b]{0.3\linewidth}
  \centering
  \centerline{\includegraphics[width=2cm]{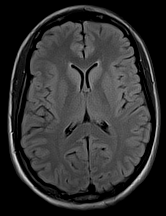}}
  \centerline{\footnotesize (c) 22.6/.92/.85/.77}\medskip
\end{minipage}
\newline
\begin{minipage}[b]{0.3\linewidth}
  \centering
  \centerline{\includegraphics[width=2cm]{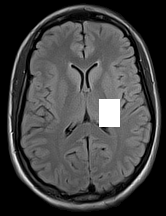}}
  \centerline{\footnotesize (d) \textbf{22.6/.98/.66/.54}}
\end{minipage}
\hfill
\begin{minipage}[b]{0.3\linewidth}
  \centering
  \centerline{\includegraphics[width=2cm]{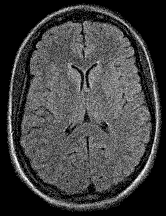}}
  \centerline{\footnotesize (e) 22.6/.64/.69/.58}
\end{minipage}
\hfill
\begin{minipage}[b]{0.3\linewidth}
  \centering
  \centerline{\includegraphics[width=2cm]{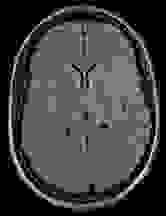}}
  \centerline{\footnotesize (f) 22.6/.63/.36/.24}
\end{minipage}
    \caption{\small A comparison of PSNR/SSIM/HaarPSI/HaarPSI$_{MED}$ for a reference MRI (a) and synthetic degradations; contrast (b), brightness (c), hole (d), white noise (e), jpeg compression (f). PSNR and SSIM fail to penalize the information loss in (d) accordingly.
    }
    \label{fig:toyexp}
\end{figure}

In Figure \ref{fig:optimization} plots of the SRCC surfaces illustrate the behavior of HaarPSI when varying $C$ and $\alpha$. These surface plots of the natural image data sets are nearly constant, while the SRCC of the medical image data sets vary greatly over the computed parameter range. A comparison of the SRCC achieved when employing the default versus optimized parameter settings is presented in Table \ref{tab:optimization}. 
For the optimized parameters (individually and combined), the SRCC of the medical data sets increases by up to $0.02$, and for the natural color images it changes negligibly when using the combined novel parameter setting instead of the default. Further, we find that the optimized parameters are similar for both medical tasks and denote HaarPSI with the newly identified parameter setting as HaarPSI$_{MED}$, optimized over both data sets. In Table \ref{tab:results}, we compare it to other common FR-IQA measures by computing the SRCC/KRCC between the manual expert annotations and the measures' evaluation values, showing that HaarPSI$_{MED}$ outperforms all other measures (including HaarPSI) on the medical data sets significantly ($p<0.05$). 
We further add the independent CT test data set to the comparison, including the measures' runtimes on the data set and approximate memory usage when evaluating one image (median over a subset of $100$ images). Lastly, we compare in Figure \ref{fig:CXR} and \ref{fig:toyexp} degraded CXR and MRI images and observe that HaarPSI$_{MED}$ qualitatively outperforms PSNR, SSIM and HaarPSI.

\begin{table}[t]
    \centering
    \resizebox{0.85\columnwidth}{!}{%
    \begin{tabular}{l |c |c| c c }
    \hline
    & \textbf{HaarPSI} & \textbf{HaarPSI$_{MED}$} & \multicolumn{2}{c}{Optimized parameter} \\  
    \hline
        & default \cite{Reisenhofer18} & CXR \& PA & CXR & PA \\
        \hline
        $C$ & 30 & 5 & 5 & 5 \\
        $\alpha$ & 4.2 & 4.9 & 4.9 & 6.3 \\
        \hline
         CXR            & 0.8680            & \textbf{0.8747}            & \textbf{0.8747} & 0.8671\\
         PA             & 0.8120            & 0.8333            & 0.8333 & \textbf{0.8336}\\
         LIVE           & \textbf{0.9588}   & 0.9569            & 0.9569 & 0.9568 \\
         LIVE$_{M}$     & 0.8861   & 0.8861            & 0.8861 & \textbf{0.8903} \\
         \hline
    \end{tabular}
    }
    \caption{\small SRCC values for the data sets CXR, PA, LIVE and LIVE$_{M}$ for default and adapted parameters $C$ and $\alpha$.}
    \label{tab:optimization}
\end{table}

\begin{table}[h]
\centering
    \resizebox{1\columnwidth}{!}{%
\begin{tabular}{c c c | c c c | c c}
\hline
  & \multicolumn{2}{c}{Natural Images} & \multicolumn{3}{c}{Medical Images} & Runtime&Memory\\
\hline
                 & LIVE             & LIVE$_{M}$        & PA                & CXR & CT & CT (\textit{s}) &CT (\textit{MB})\\
\hline
PSNR            & .90/.73 \pgood    & .63/.46 \pgood    & .71/.54  \pgood   & .52/.37  \pgood   &  .93/.78 \pgood   & 0.1  & 860 \\
SSIM            & .89/.72 \pgood    & .50/.35 \pgood    & .62/.49  \pgood   & .78/.57  \pgood   &  .92/.77 \pgood   & 0.9  & 874 \\ 
HaarPSI$_{MED}$ & \textbf{.96/.82}           & \textbf{.89/.70}           & \textbf{.83/.68}  & \textbf{.87/.67}  &  \textbf{.95/.81} & 1.1  & 454 \\
HaarPSI         & \textbf{.96/.83}           & \textbf{.89/.70}           & .81/.65  \pgood   & \textbf{.87/.66}  \pgood   &  \pgood .94/.80   & 1.3  & 454 \\
MS-SSIM         & .94/.79 \pgood    & .78/.59  \pgood   & .81/.65  \pgood   & .84/.63  \pgood   &  .93/.79 \pgood   & 0.5  & 873 \\
IW-SSIM         & -                 & -                 & .79/.63  \pgood   & .80/.59  \pgood   &  .92/.76 \pgood   & 9.4  & 962 \\
GMSD            & -                 & -                 & .77/.60  \pgood   & .86/.65  \pgood   &  .92/.77 \pgood   & 0.3  & 963 \\
FSIM            & \textbf{.96/.84} \pbad & .87/.69 \pgood & .78/.62  \pgood   & .83/.61  \pgood   &  \textbf{.95/.81}          & 9.3 &963\\
MDSI            & \textbf{.96/.82}     & .87/.68\pgood     & .68/.51  \pgood   & .83/.62  \pgood   &  .94/.79 \pgood   & 0.7  & 969\\
LPIPS$_{Alex}$  & .93/.79  \pgood   & .78/.59  \pgood   & .76/.59  \pgood   & .84/.63  \pgood   &  .86/.69 \pgood   &  4.8 &899\\
DISTS           & .95/.81  \pgood   & .81/.62  \pgood   & .70/.56  \pgood   & .82/.60  \pgood   &  .84/.65 \pgood   & 178.5 & 851\\
\hline
\end{tabular} 
}
\caption{\small Absolute SRCC/KRCC values between a measure's evaluation value and the manual quality ratings. The highest SRCC values have been printed in bold for each data set. 
For the CT data we also provide a measure`s runtime for the whole test data set (272 images) and the MB usage per image (median over 100 images). Green/red coloring indicates that the HaarPSI$_{MED}$ SRCC or KRCC is significantly better/worse with a significance level $p=0.05$. }
\label{tab:results}
\end{table}

\section{Discussion}
In Figure \ref{fig:optimization}, we see that the medical image data sets are more sensitive to parameter variation in HaarPSI compared to the natural images. Although PA and CXR comprise very different modalities and were rated for different tasks, they show related behavior. The best results are achieved for a lower parameter $C$, corresponding to increased sensitivity to high frequencies (cf.~\cite{Reisenhofer18}), while for the natural image data sets the results are less influenced by the choice of $C$. 
In our experiments, the parameter $\alpha$ only strongly influences the results of the CXR data set. 
It is noteworthy that regarding overall generalizability it is here more beneficial to adjust the parameters accurately to the medical images, corresponding to increased sensitivity to details, than to use the default parameters based on natural images. This may relate to the complexity of tasks, which is reflected by LIVE$_{M}$ being more sensitive to parameter changes than LIVE, as it includes multiple instead of just single distortions at once.
The correlation coefficients in Table \ref{tab:optimization} confirm the described observations. We additionally observe that optimizing independently for PA or CXR also leads to a suitable parameter choice for all employed data sets. However, to avoid substantial overfitting, HaarPSI$_{MED}$ is chosen as the parameter setting corresponding to the optimization of the mean SRCC obtained by PA and CXR.  

In Table \ref{tab:results} we include the independent CT test data set and compare the results to other commonly used FR-IQA measures. 
We observe that HaarPSI, FSIM and MDSI are top performers for the natural images, whereas HaarPSI$_{MED}$ provides the highest SRCC/KRCC for all medical image data sets with statistical significance ($p<0.05$), whilst keeping high results for the natural images. Furthermore, HaarPSI and HaarPSI$_{MED}$ are relatively fast while utilizing the least memory. However, it should be noted that HaarPSI, HaarPSI$_{MED}$ and LPIPS are implemented in Python, whilst all other measures are originally written in Matlab, which affects the memory usage, see S.M. section 3 for further details and an analysis on all employed data sets. Lastly, in a qualitative analysis of examples from the CXR data set in Figure \ref{fig:CXR}, we observe that HaarPSI and HaarPSI$_{MED}$ perform better than the FR-IQA measures PSNR and SSIM, all of which wrongly rank (b) better than (c). Compared to HaarPSI, HaarPSI$_{MED}$ is able to differentiate more clearly between (b) and (c). To further investigate the suitability of HaarPSI$_{MED}$ for unrelated data sets, we employ MRI brain images with synthetic toy degradations, see Figure \ref{fig:toyexp}. In these images, many IQA measures, including PSNR and SSIM, struggle to detect local information loss, cf.~\cite{breger2024study1}. Compared to HaarPSI, the novel parameter setting HaarPSI$_{MED}$ further improves the penalization of local deterioration. It successfully identifies this type of distortion, even though it is not optimized for MRI images, demonstrating promising behavior for generalizability to other medical tasks.

In this study, we have seen that a careful choice of parameters for complex tasks, such as medical images, is of utmost importance to ensure optimal suitability of HaarPSI. In order to encourage a careful choice of parameters, our PyTorch implementation\footnote{\url{https://github.com/ideal-iqa/haarpsi-pytorch}} of HaarPSI requires a compulsory parameter choice. The parameter settings of HaarPSI and HaarPSI$_{MED}$ are provided. 
\vspace{-0.3cm}
\section{Limitations}
HaarPSI shows promising behavior for a generalizable FR-IQA measures, that is suitable to assess images across domains. Nevertheless, that kind of assessment cannot replace the choice of evaluation methods that have been developed for certain image classes and tasks at hand. Whilst generalizability is one important feature, attention to specific needed details is critical in highly complex tasks. 
In order to confirm the observed trend of parameter choices in HaarPSI for medical images, further analysis with medical data is desirable. For such tasks the lack of annotated data sets is still hindering comprehensive research. In addition to overcoming that, self-supervised optimization schemes might provide a promising research direction. 
\vspace{-0.4cm}
\section{Conclusion}
We presented an adaption of the FR-IQA measure HaarPSI to the medical domain based on two annotated medical image data sets. 
The precise parameter choices show a stronger impact on the medical images' evaluation compared to two tested natural image data sets. We demonstrate that the adjustment through parameter optimization on rated photoacoustic and chest X-ray images yields significant ($p<0.05$) improvement of the correlation to manual expert ratings for the employed data as well as on an independent CT data set.  
HaarPSI$_{MED}$ also clearly outperforms other common FR-IQA measures for the medical data sets whilst generalizing well to the natural images. In visual examples, HaarPSI$_{MED}$ demonstrates greater sensitivity than HaarPSI to distinct quality loss.

This study shows the great potential of parameter adaptation of the FR-IQA measure HaarPSI to medical images. A PyTorch implementation of HaarPSI$_{MED}$ with the novel parameter choices is provided on GitHub.
\section{Ethical standards}
\label{sec:ethics}

The employed data sets LIVE \cite{live1,live2_ssim,live3}, LIVE$_M$ \cite{livemulti1, livemulti2}, PA \cite{Janek7, breger2024study2} and CT \cite{MICCAI2023} are publicly available. The CXR data has been acquired following the ethical approval under IRAS number 282705 and is in progress to be made available in accordance with the ethical agreements. 

\section{Acknowledgments}
The authors wish to acknowledge support from the EU/EFPIA Innovative Medicines Initiative 2 Joint Undertaking - DRAGON (101005122) (A.Br., I.S., M.R., J.Ba., C.B.S.); the Austrian Science Fund (FWF) through project T1307 (A.Br., C.K.); the German Research Foundation through the grant GR 5824/1 (J.G.); the National Institute for Health and Care Research (NIHR) Cambridge Biomedical Research Centre (BRC-1215-20014) (I.S.); the Wellcome Trust (J.H.F.R.); the British Heart Foundation (J.H.F.R.); the NIHR Cambridge Biomedical Research Centre (J.H.F.R.); the EPSRC Cambridge Mathematics of Information in Healthcare Hub EP/T017961/1 (M.R., J.H.F.R., C.B.S.); and the Trinity Challenge BloodCounts! project (M.R., C.B.S.). 

Please note that the content of this publication reflects the authors’ views and that neither NIHR, the Department of Health and Social Care, IMI, the European Union, EFPIA, nor the DRAGON consortium are responsible for any use that may be made of the information contained therein. 
\small
\bibliographystyle{IEEEbib}
\bibliography{strings,refs_et_al_nameshortened}

\begin{thebibliography}{10}

\bibitem{639240}
B.~Girod,
\newblock ``Psychovisual aspects of image processing: What's wrong with mean
  squared error?,''
\newblock in {\em Proceedings of the Seventh Workshop on Multidimensional
  Signal Processing}, 1991, pp. P.2--P.2.

\bibitem{live2_ssim}
Z.~Wang, A.C. Bovik, et~al.,
\newblock ``Image quality assessment: from error visibility to structural
  similarity,''
\newblock {\em IEEE Trans. Image Process.}, vol. 13, no. 4, pp. 600--612, 2004.

\bibitem{breger2024study1}
A.~Breger, A.~Biguri, et~al.,
\newblock ``A study of why we need to reassess full reference image quality
  assessment with medical images,''
\newblock {\em Journal of Imaging Informatics in Medicine}, Mar 2025.

\bibitem{iwssim}
Z.~Wang and Q.~Li,
\newblock ``Information content weighting for perceptual image quality
  assessment,''
\newblock {\em IEEE Trans. Image Process.}, vol. 20, no. 5, pp. 1185--1198,
  2011.

\bibitem{msssim}
E.~P.~Simoncelli Z.~Wang and A.~C. Bovik,
\newblock ``Multiscale structural similarity for image quality assessment,''
\newblock in {\em 37th Asilomar Conference on Signals, Systems \& Computers},
  2003, vol.~2, pp. 1398--1402 Vol.2.

\bibitem{lpips}
R.~Zhang, P.~Isola, et~al.,
\newblock ``The unreasonable effectiveness of deep features as a perceptual
  metric,''
\newblock in {\em 2018 IEEE/CVF CVPR}, 2018, pp. 586--595.

\bibitem{9298952}
K.~Ding, K.~Ma, et~al.,
\newblock ``Image quality assessment: Unifying structure and texture
  similarity,''
\newblock {\em IEEE Trans. Pattern Anal. Mach. Intell.}, vol. 44, no. 5, pp.
  2567--2581, 2022.

\bibitem{breger2024study2}
A.~Breger, C.~Karner, et~al.,
\newblock ``A study on the adequacy of common iqa measures for medical
  images,''
\newblock in {\em Proceedings of MICAD}, 2024, Springer LNEE.

\bibitem{IQAdylovmri}
S.~Kastryulin, J.~Zakirov, et~al.,
\newblock ``Image quality assessment for magnetic resonance imaging,''
\newblock {\em IEEE Access}, vol. 11, pp. 14154--14168, 2023.

\bibitem{Reisenhofer18}
R.~Reisenhofer, S.~Bosse, et~al.,
\newblock ``A haar wavelet-based perceptual similarity index for image quality
  assessment,''
\newblock {\em Signal Process. Image Commun.}, vol. 61, pp. 33--43, 2018.

\bibitem{live1}
H.~R. Sheikh, Z.~Wang, et~al.,
\newblock ``Live image quality assessment database release 2,''
  http://live.ece.utexas.edu/research/quality.

\bibitem{live3}
H.R. Sheikh, M.F. Sabir, and A.C. Bovik,
\newblock ``A statistical evaluation of recent full reference image quality
  assessment algorithms,''
\newblock {\em IEEE Trans. Image Process.}, vol. 15, no. 11, pp. 3440--3451,
  2006.

\bibitem{tid2008}
N.~Ponomarenko, V.~Lukin, et~al.,
\newblock ``Tid2008-a database for evaluation of full-reference visual quality
  assessment metrics,''
\newblock {\em Advances of modern radioelectronics}, vol. 10, no. 4, pp.
  30--45, 2009.

\bibitem{tid2013}
N.~Ponomarenko, L.~Jin, et~al.,
\newblock ``Image database tid2013: Peculiarities, results and perspectives,''
\newblock {\em Signal Process. Image Commun.}, vol. 30, pp. 57--77, 2015.

\bibitem{csiq}
E.C. Larson and D.~M. Chandler,
\newblock ``{Most apparent distortion: full-reference image quality assessment
  and the role of strategy},''
\newblock {\em J. Electron. Imaging}, vol. 19, no. 1, pp. 011006, 2010.

\bibitem{Janek7}
J.~Gr\"ohl, T.~Else, et~al.,
\newblock ``Moving beyond simulation: Data-driven quantitative photoacoustic
  imaging using tissue-mimicking phantoms,''
\newblock {\em IEEE Trans Med Imaging}, vol. 43, no. 3, pp. 1214--1224, 2024.

\bibitem{photoacoustic_dataset}
A.~Breger, J.~Gröhl, and T.~Else,
\newblock ``Photoacoustic data annotations supplementing the paper: "a study on
  the adequacy of common iqa measures for medical images",''
  https://doi.org/10.5281/zenodo.13325196.

\bibitem{MICCAI2023}
W.~Lee, F.~Wagner, et~al.,
\newblock ``Low-dose computed tomography perceptual image quality assessment
  grand challenge dataset (miccai 2023),''
  https://doi.org/10.5281/zenodo.7833096.

\bibitem{gmsd}
W.~Xue, L.~Zhang, et~al.,
\newblock ``Gradient magnitude similarity deviation: A highly efficient
  perceptual image quality index,''
\newblock {\em IEEE Trans. Image Process.}, vol. 23, no. 2, pp. 684--695, 2014.

\bibitem{fsim}
L.~Zhang, L.~Zhang, et~al.,
\newblock ``Fsim: a feature similarity index for image quality assessment.,''
\newblock {\em IEEE Trans. Image Process.}, vol. 20, no. 8, pp. 2378--2386, Aug
  2011.

\bibitem{mdsi}
H.~Z.~Nafchi, A.~Shahkolaei, et~al.,
\newblock ``Mean deviation similarity index: Efficient and reliable
  full-reference image quality evaluator,''
\newblock {\em IEEE Access}, vol. 4, pp. 5579--5590, 2016.

\bibitem{livemulti2}
D.~Jayaraman, A.~Mittal, et~al.,
\newblock ``Objective quality assessment of multiply distorted images,''
\newblock in {\em 2012 Conference Record of the 46th Asilomar Conference on
  Signals, Systems \& Computers}, 2012, pp. 1693--1697.

\bibitem{livemulti1}
``Live subjective multiply distorted image quality database,''
  https://live.ece.utexas.edu/research/Quality/ \\
  live\_multidistortedimage.html.

\end{thebibliography}


\begin{thebibliography}{10}

\bibitem{srcc}
E.~C. Fieller, H.~O. Hartley, and E.~S. Pearson,
\newblock ``{Tests for rank correlation coefficients. I},''
\newblock {\em Biometrika}, vol. 44, no. 3-4, pp. 470--481, 12 1957.

\bibitem{krcc}
M.~G. Kendall,
\newblock ``{A new measure of rank correlation},''
\newblock {\em Biometrika}, vol. 30, no. 1-2, pp. 81--93, 06 1938.

\bibitem{correlation_overview}
Harry Khamis,
\newblock ``Measures of association: How to choose?,''
\newblock {\em Journal of Diagnostic Medical Sonography}, vol. 24, no. 3, pp.
  155--162, 2008.

\bibitem{live1}
H.~R. Sheikh, Z.~Wang, et~al.,
\newblock ``Live image quality assessment database release 2,''
  http://live.ece.utexas.edu/research/quality.

\bibitem{live2_ssim}
Z.~Wang, A.C. Bovik, et~al.,
\newblock ``Image quality assessment: from error visibility to structural
  similarity,''
\newblock {\em IEEE Trans. Image Process.}, vol. 13, no. 4, pp. 600--612, 2004.

\bibitem{live3}
H.R. Sheikh, M.F. Sabir, and A.C. Bovik,
\newblock ``A statistical evaluation of recent full reference image quality
  assessment algorithms,''
\newblock {\em IEEE Trans. Image Process.}, vol. 15, no. 11, pp. 3440--3451,
  2006.

\bibitem{livemulti1}
``Live subjective multiply distorted image quality database,''
  https://live.ece.utexas.edu/research/Quality/ \\
  live\_multidistortedimage.html.

\bibitem{livemulti2}
D.~Jayaraman, A.~Mittal, et~al.,
\newblock ``Objective quality assessment of multiply distorted images,''
\newblock in {\em 2012 Conference Record of the 46th Asilomar Conference on
  Signals, Systems \& Computers}, 2012, pp. 1693--1697.

\bibitem{breger2024study2}
A.~Breger, C.~Karner, et~al.,
\newblock ``A study on the adequacy of common iqa measures for medical
  images,''
\newblock in {\em Proceedings of MICAD}, 2024, Springer LNEE.

\bibitem{speedyiqa}
Ian Selby,
\newblock ``Github repository speedyiqa,''
  https://github.com/selbs/speedy\_iqa.

\bibitem{Reisenhofer18}
R.~Reisenhofer, S.~Bosse, et~al.,
\newblock ``A haar wavelet-based perceptual similarity index for image quality
  assessment,''
\newblock {\em Signal Process. Image Commun.}, vol. 61, pp. 33--43, 2018.

\end{thebibliography}

\end{document}


\maketitle
This supplemental material complements the main paper by providing further insights. It contains in Section \ref{sec1} the explicit definition of the Spearman Rank Correlation Coefficient (SRCC) \cite{srcc} and the Kendall Rank Correlation Coefficient (KRCC) \cite{krcc} that are employed to evaluate the correlation between the expert's ratings and the IQA measures' output. In Section \ref{sec2} we compare the results of the conducted experiments when adapting three of the data sets, namely the two natural image data sets transformed to the grayscale color space (to match the target domain of the medical images) and the CT data without any cropping to the region of interest (ROI). Finally, in Section \ref{secfinal}, we provide a detailed runtime and memory usage analysis comparing HaarPSI$_{MED}$ with the other employed measures on all data sets that are part of the study.
 \vspace{-0.5cm}
\section{Correlation Coefficients}\label{sec1}
To assess the alignment of the z-scores of the manual ratings, and the IQA measure's evaluations, we employ the SRCC and KRCC, see e.g.~\cite{correlation_overview}. In contrast to the Pearson Correlation Coefficient, which computes the linear relationship of two variables, the SRCC and KRCC compute how strongly the ranks of the entries of a vector $x\in \mathbb{R}^n$ (here, containing the mean z-scored image ratings by the graders) and $y\in \mathbb{R}^n$ (here, containing the image evaluation values provided by an IQA measure), correlate.  \\
The SRCC can be written as 
$$
\SRCC(x,y)=1-\frac{6 \sum_{i=1}^{n} (\mathrm{R}(x_i)-\mathrm{R}(y_i))^ 2}{n(n^ 2-1)},
$$
where $\mathrm{R}(x_i)$ denotes the rank of the $i$-th entry of the vector $x$, corresponding here to the rating of the $i$-th image. \\The KRCC can be formulated as 
\begin{align*}
\KRCC(x,y)&=\frac{2\sum_{i=1}^{n-1}\sum_{j=i+1}^{n}\xi(x_i,x_j,y_i,y_j)}{n(n-1)}, \\ \text{with } 
\xi(x_i,x_j,y_i,y_j)&=\begin{cases}
      1 & \text{for} (x_i-x_j)(y_i-y_j)>0,\\
      0 &  \text{for} (x_i-x_j)(y_i-y_j)=0,\\
      -1 & \text{for} (x_i-x_j)(y_i-y_j)<0. \\
\end{cases}
\end{align*}
\section{Additional experiments}\label{sec2}
\subsection{Data}
The employed natural images in the main study, namely LIVE (\textit{Laboratory for Image \& Video Engineering Image Quality Assessment Database Release 2}~\cite{live1,live2_ssim,live3}) and LIVE{$_M$} (\textit{Laboratory for Image \& Video Engineering Multiply Distorted Image Quality Database}~\cite{livemulti1, livemulti2}) are originally RGB color images. This is a fundamental difference to the employed medical grayscale images in terms of the color space. To better understand the influence of the parameter choices in HaarPSI regarding the content rather than color space, we have additionally converted the natural color images to grayscale with the built-in MATLAB function \textit{rgb2gray} which computes for a color image $I^c \in \mathbb{R}^{m \times n \times 3}$ with $m,n \in \mathbb{N}$ the grayscale image $I \in \mathbb{R}^{m \times n}$ by
$$
I_{i,j}=0.2989 \cdot I^c_{i,j,1} + 0.5870 \cdot I^c_{i,j,2} + 0.1140 \cdot I^c_{i,j,3},
$$
where $i \in \{1, \ldots, m\}$ and  $j \in \{1, \ldots, n\}$ denote the location of a pixel in the image. After the conversion to grayscale, we computed the normalized grayscale images denoted as $\hat{I} \in \mathbb{R}^{m \times n}$ using the built-in MATLAB function \textit{mat2gray} by
$$
\hat{I}_{i,j}=\frac{I_{i,j}-\min_{i,j}(I_{i,j})}{\max_{i,j}(I_{i,j})-\min_{i,j}(I_{i,j})}.
$$
Image quality annotations for the grayscale data sets were obtained by $5$ volunteers in 2024 \cite{breger2024study2} using the SpeedyIQA application \cite{speedyiqa}. The annotators rated the images as 1 (very poor), 2 (poor), 3 (good) or 4 (very good) regarding the ability to identify the detailed image content in comparison to the given reference image. The annotations have been made available on GitHub (\url{https://github.com/ideal-iqa/iqa-eval}).  In Table \ref{tab:annotation_correlation} we compute the correlation between the grayscale and color annotations, illustrating a different rating behavior for the grayscale images. This may be caused by the different population of volunteers, but may also be due the visualization in a different color space. 

We denote the two grayscale natural image data sets as LIVE$^*$ (982 grayscale images, up to $768 \times 512$ pixels, ratings by 5 volunteers) and LIVE$_{M}^*$ (405 grayscale images, up to $1280 \times 720$ pixels, ratings by 5 volunteers).  

Furthermore, we study the uncropped \textit{Low-dose Computed Tomography Perceptual Image Quality Assessment Grand Challenge data set} (CT$_{full}$, $272$ test grayscale images, $512 \times 512$ pixels, $6$ expert ratings), to illustrate the effect of cropping to the ROI, which is approximately $70$\% of the original image, in the main paper. An example comparing an image from the CT and CT$_{full}$ data set is presented in Figure \ref{fig:CT_crop_example}.
\vspace{-0.2cm}
\subsection{Results and Discussion}
\begin{figure}[h]
\begin{minipage}[b]{1\linewidth}
  \centering
  \centerline{\includegraphics[width=4cm]{./images/corrmat_cxr_photoacoustic_colorbar_new.png}}
  \centerline{\small(a) Mean SRCC of CXR and PA}\medskip
\end{minipage}
\begin{minipage}[b]{0.48\linewidth}
  \centering
\centerline{\includegraphics[width=4.0cm]{./images/corrmat_live_color_new.png}}
  \centerline{(b) LIVE}\medskip
\end{minipage}
\begin{minipage}[b]{0.48\linewidth}
  \centering
\centerline{\includegraphics[width=4.0cm]{./images/corrmat_live_multi_color_new.png}}
  \centerline{(c) LIVE$_M$}\medskip
\end{minipage}
\begin{minipage}[b]{0.48\linewidth}
  \centering
\centerline{\includegraphics[width=4.0cm]{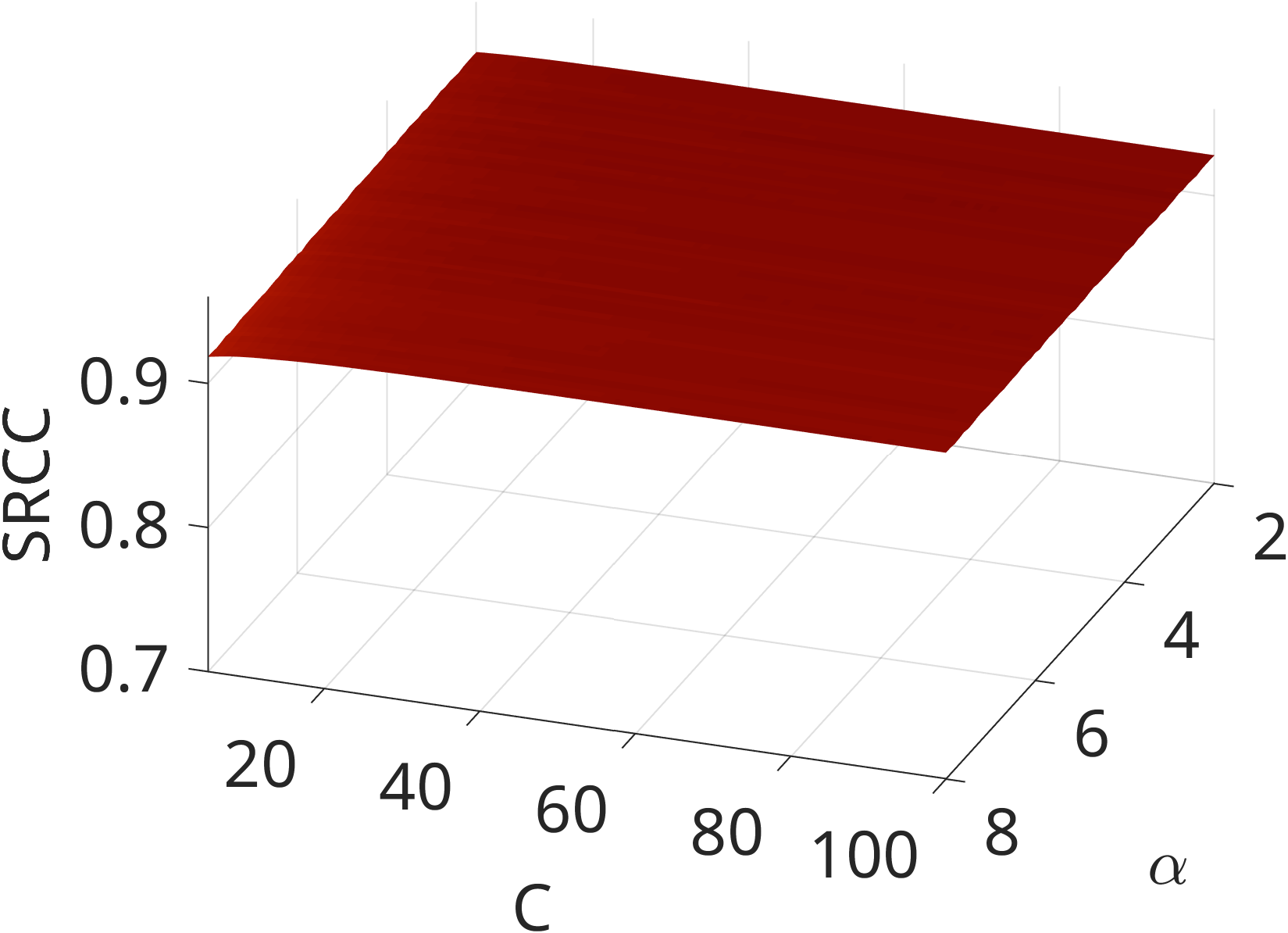}}
  \centerline{(d) LIVE$^*$}\medskip
\end{minipage}
\hfill
\begin{minipage}[b]{0.48\linewidth}
  \centering
  \centerline{\includegraphics[width=4.0cm]{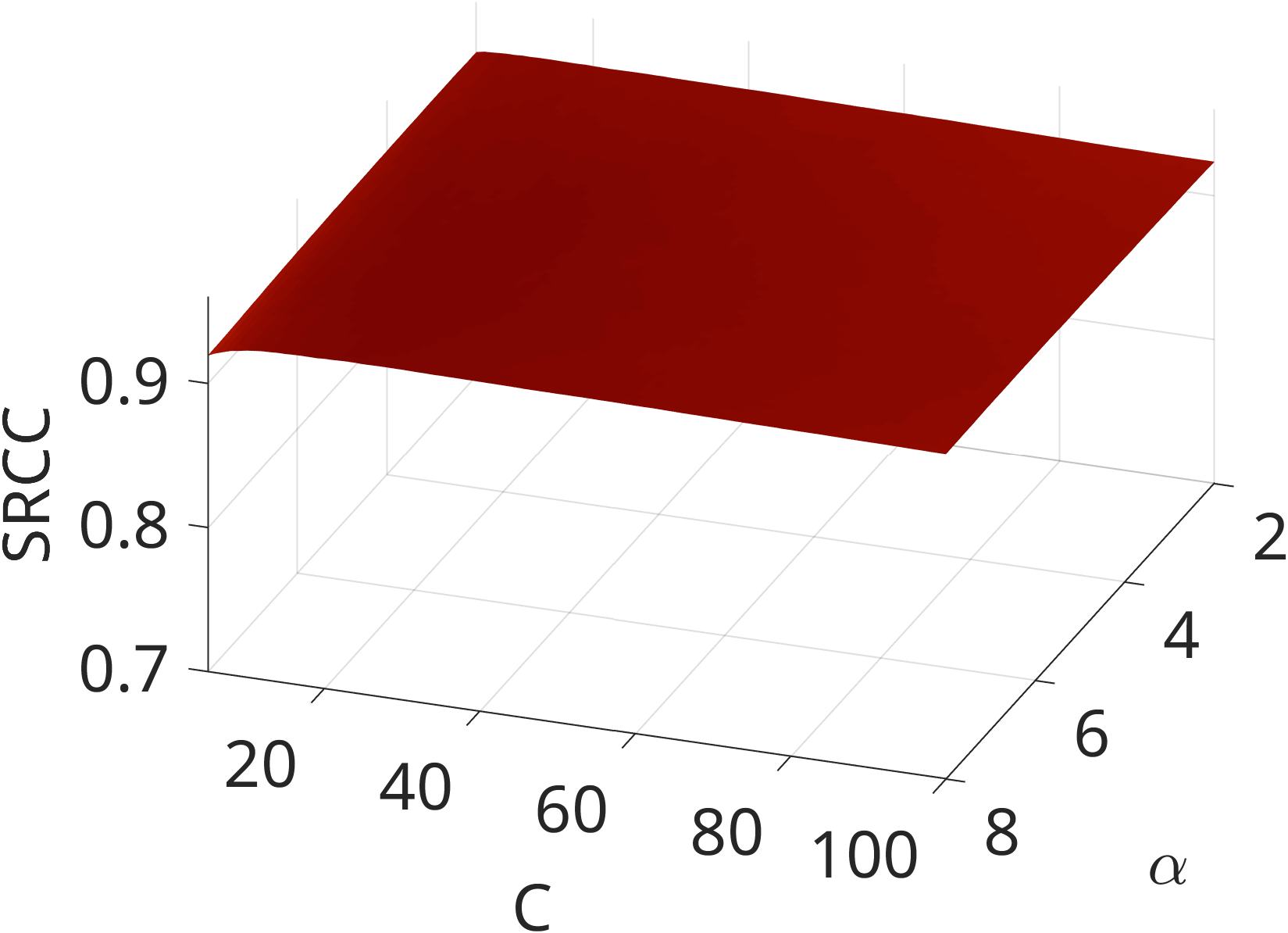}}
  \centerline{(e) LIVE$^*_{M}$}\medskip
\end{minipage}
\caption{SRCC values for HaarPSI evaluation of several annotated data sets for the HaarPSI parameters $C$ in $\{5, 6, \ldots, 100\}$ and $\alpha$ in $\{2,2.1, \ldots, 8\}$.}
\label{fig:optimization_supplementary}
\end{figure}

In Figure \ref{fig:optimization_supplementary} plots of the SRCC surfaces illustrate the behavior of HaarPSI as the parameters $C$ and $\alpha$ vary, and the results for the parameter settings HaarPSI$_{MED}$ are stated in Table \ref{tab:results_supplementary}. Similarly to the observations in \cite{Reisenhofer18}, we see in Figure \ref{fig:optimization_supplementary} a near linear dependence of the SRCC to $C$ and $\alpha$. Both, high $C$ as well as low $\alpha$ values, decrease the sensitivity of HaarPSI to high-frequency components and we can follow that the optimal parameters for the medical image data sets foster increased sensitivity to details. We observe that both grayscale data sets (LIVE$^*$ and LIVE$_{M}^*$) are sensitive to parameter changes in HaarPSI, especially in the parameter $C$ and show a slight performance drop with lower $C$. This is in conflict with the better performance of lower $C$ values for the tested medical data sets and therefore leads to slightly lower results with HaarPSI$_{MED}$, being statistically significant ($p<0.05$) only on the grayscale data set LIVE$^*$. These results differ from the original color image data sets, which might be due to the different annotations (see Table \ref{tab:annotation_correlation}) or because of varying behavior of the IQA measures depending on the color space (see e.g.~\cite{Reisenhofer18}). We further observe that the top performers on LIVE$^*$ are HaarPSI, FSIM and on LIVE$_M^*$ are HaarPSI, IW-SSIM, FSIM and MDSI. Interestingly, most measures hold higher correlations to the color image ratings for the LIVE data set, but the opposite holds for the LIVE$_{M}$ data set. 

\begin{figure}[h]
\begin{minipage}[b]{0.48\linewidth}
  \centering
\centerline{\includegraphics[width=4.0cm]{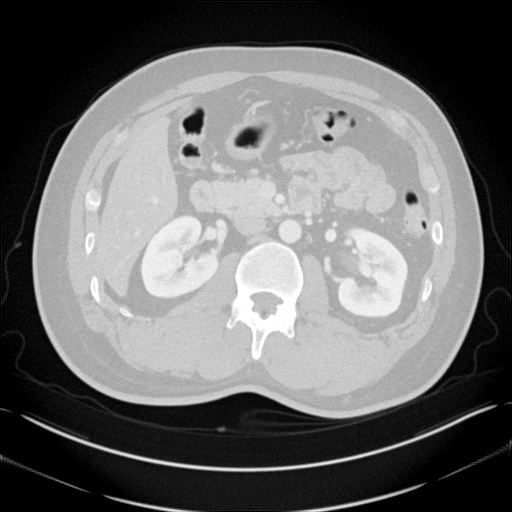}}
  \caption*{CT$_{full}$: $512 \times 512$ pixels \newline PSNR/SSIM: 38.9/0.93}\medskip
\end{minipage}
\hfill
\begin{minipage}[b]{0.48\linewidth}
  \centering
  \centerline{\includegraphics[width=4.0cm]{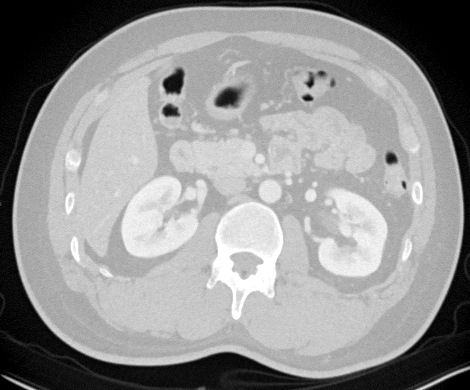}}
  \caption*{CT: $470 \times 390$ pixels\newline PSNR/SSIM: 38.2/0.92}\medskip
\end{minipage}
\caption{Comparison of the full CT image (left) and the cropped version (right), visualizing the impact of cropping $30$\% of the image to reduce the number of background pixels.}
\label{fig:CT_crop_example}
\end{figure}

Moreover, in Table \ref{tab:results_supplementary} we compare the results of the IQA measures on the CT$_{full}$ versus the cropped CT data set, reducing the background pixels outside the ROI. The background, about $30$\% of the uncropped image, has a huge influence due to the background noise unrelated to the ROI which is reflected in the lower SRCC/KRCC values for the uncropped images in comparison to the cropped ones. Interestingly, the PSNR, HaarPSI and the SSIM based IQA measures all yield nearly the same correlation coefficients for the uncropped images, resulting in a much smaller variation between the measures than for the cropped versions. This emphasizes that the huge amount of background pixels is compromising the focus on significant information. In Fig. \ref{fig:CT_crop_example} a cropped and a corresponding uncropped example image is shown together with the IQA scores of PSNR and SSIM, yielding lower values for the uncropped version. These small changes add up over the whole data set to the overall improved SRCC/KRCC scores for the cropped images. We therefore conclude that cropping the data to the ROI helps several IQA measures to focus on the relevant information.  

\begin{table*}[h]
\centering
    \resizebox{1.6\columnwidth}{!}{%
\begin{tabular}{c c c c c | c c c  c}
\hline
  & \multicolumn{4}{c}{Natural Images} & \multicolumn{4}{c}{Medical Images} \\
\hline
                & LIVE$^*$           & LIVE     &  LIVE$^*_{M}$      & LIVE$_{M}$ & PA        & CXR & CT & CT$_{full}$\\
\hline
PSNR            & .88/.71 \pgood    & .90/.73  \pgood   & .75/.57  \pgood  & .63/.46  \pgood    & .71/.54 \pgood   & .52/.37 \pgood &   .93/.78 \pgood & .92/.77\\
SSIM            & .88/.72 \pgood    & .89/.72  \pgood   & .67/.50  \pgood  & .50/.35  \pgood    & .62/.49 \pgood & .78/.57 \pgood &   .92/.77 \pgood& .92/.77\\ 
HaarPSI$_{MED}$ & .92/.78           & \textbf{.96/.82}     & .92/.76          & \textbf{.89/.70}      & \textbf{.83/.68}& \textbf{.87/.67}    &   \textbf{.95/.81} & .92/.77\\
HaarPSI         & \textbf{.93/.79}  \pbad    & \textbf{.96/.83}     & \textbf{.93/.77}          & \textbf{.89/.70}     & .81/.65 \pgood  & \textbf{.87/.66}  \pgood &   .94/.80 \pgood & .92/.76 \pgood \\
MS-SSIM         & .91/.77  \pgood   & .94/.79  \pgood   & .89/.71 \pgood   & .78/.59  \pgood    & .81/.65 \pgood  & .84/.63 \pgood &    .93/.79 \pgood& .92/.77\\
IW-SSIM         & .92/.79           & -           & \textbf{.93/.78} \pbad & -             & .79/.63  \pgood  & .80/.59 \pgood &   .92/.76 \pgood & .92/.77\\
GMSD            & .92/.79           & -           & .92/.75          & -            & .77/.60   \pgood   & .86/.65   \pgood &   .92/.77\pgood & .89/.72 \pgood \\
FSIM            & \textbf{.93/.80} \pbad& \textbf{.96/.84} \pbad & \textbf{.93/.77}      & .87/.69   \pgood   & .78/.62 \pgood   & .83/.61 \pgood  &   \textbf{.95/.81} & \textbf{.94/.79} \pbad\\
MDSI            & .92/.78           & \textbf{.96/.82}    & \textbf{.93/.77}  \pbad   & .87/.68  \pgood & .68/.51 \pgood  & .83/.62 \pgood  &  .94/.79  \pgood& .91/.74 \pgood\\
LPIPS$_{Alex}$  & .90/.75 \pgood    & .93/.79 \pgood    & .77/.59 \pgood   & .78/.59   \pgood   & .76/.59  \pgood  & .84/.63 \pgood &   .86/.69  \pgood& .88/.71 \pgood \\
DISTS           & .91/.76 \pgood    & .95/.81 \pgood     & .74/.56 \pgood   & .81/.62  \pgood    & .70/.56 \pgood  & .82/.60 \pgood  &   .84/.65\pgood & .87/.70 \pgood \\
\hline
\end{tabular} 
}
\caption{Absolute SRCC/KRCC values of the measure's evaluation value and the z-score of manual quality ratings. The highest SRCC values have been printed in bold for each data set. Both IW-SSIM and GMSD only evaluate grayscale images. Green/red coloring indicates that the HaarPSI$_{MED}$ SRCC or KRCC is significantly better/worse with a significance level $p=0.05$.}
\label{tab:results_supplementary}
\end{table*}

\begin{table}[h]
    \centering
    \begin{tabular}{|c|c| }
         &  LIVE\\
         \hline
        LIVE$^*$ & .92/.78\\
        \hline
    \end{tabular}
    \begin{tabular}{|c|c| }
          & LIVE$_M$\\
         \hline
        LIVE$^*_M$ &  .89/.72\\
        \hline
    
    \end{tabular}
    \caption{Absolute SRCC/KRCC values between the z-scores of the color and grayscale manual quality ratings.}
    \label{tab:annotation_correlation}
\end{table}

\vspace{-0.3cm}
\section{Computational resources}\label{secfinal}
In Table \ref{tab:runtime} we see that our PyTorch HaarPSI$_{MED}$ implementation (\url{https://github.com/ideal-iqa/haarpsi-pytorch}) offers a competitive runtime across all data sets. Note that we do not include HaarPSI in this comparison as it just differs regarding the parameters $C$ and $\alpha$ which does not impact the computational resources. PSNR is clearly the fastest algorithm due to its simplicity and correspondence to a norm, in contrast to DISTS, which clearly underperforms all other tested algorithms time-wise. Both DISTS and LPIPS$_{Alex}$ are deep learning based IQA measures, however, the number of parameters in the neural network of DISTS is approximately $6$ times the size of LPIPS$_{Alex}$ ($\sim$14.5 versus $\sim$2.5 million). Furthermore, the used DISTS implementation is written in Matlab and does not utilize an optimized neural network library, whilst LPIPS$_{Alex}$ is implemented in Python and based on a highly optimized PyTorch library. The memory analysis in Table \ref{tab:memory} shows that our PyTorch HaarPSI implementation requires by far the least memory of all algorithms on every data set.
In order to understand the impact of the programming language further, we also compare the original HaarPSI implementation in Matlab (\url{https://github.com/rgcda/haarpsi}).
As can be seen when comparing the runtime and memory usage of the Python and Matlab implementation of HaarPSI and HaarPSI$_{MED}$, the programming language of the chosen implementation indeed has a huge impact on the computational resources. The Matlab implementation runs at least $25$\% faster as the Python implementation, while at the same time it uses nearly twice as much memory. 

\small
\bibliographystyle{IEEEbib}
\bibliography{strings,refs_et_al_nameshortened}

\begin{table*}[h]
\centering
    \resizebox{\linewidth}{!}{%
\begin{tabular}{c | c c c c | c c c  c}
\hline
  & \multicolumn{4}{c}{Natural Images} & \multicolumn{4}{c}{Medical Images}\\
\hline
                & LIVE$^*$ ($982$)            & LIVE    &  LIVE$^*_{M}$ ($405$)     & LIVE$_{M}$ & PA ($1134$)       & CXR ($1571$) & CT ($272$) & CT$_{full}$ \\
\hline

PSNR            & \textbf{0.4} & \textbf{1.4} & \textbf{0.4} & \textbf{1.7} & \textbf{0.2} & \textbf{12.6} & \textbf{0.1} & \textbf{0.3} \\
SSIM & 5.9 & 18.8 & 6.0 & 36.1 & 2.2 & 193.8 & 0.9 & 5.6  \\
HaarPSI$^P_{MED}$ & 7.5 & 14.0 &13.7 & 29.7 & 3.2 & 359.2 & 1.1 & 6.1 \\
HaarPSI$_{MED}$ & 4.7 & 10.3 & 4.5 & 10.6 & 1.8 & 123.6 & 0.8 & 4.2 \\
MS-SSIM         & 1.9 & 26.9 & 1.6 & 26.5 & 1.0 & 31.5 & 0.5 & 1.9 \\
IW-SSIM         & 62.6 & - & 74.8 & - & 20.5 & 2333.9 & 9.4 & 67.0 \\
GMSD            & 1.4 & - & 1.4 & - & 0.7 & 27.9 & 0.3 & 1.4 \\
FSIM            & 49.4 & 56.8 & 28.7 & 36.0 & 46.7 & 224.3 & 9.3 & 40.6 \\
MDSI            & 4.3 & 4.5 & 4.0 & 4.3 & 3.0 & 159.7 & 0.7 & 4.1 \\
LPIPS$^P_{Alex}$  & 42.0 & 47.1 & 40.5 & 38.4 & 10.0 & 829.2 & 4.8 & 31.7 \\
DISTS           & 1231.4 & 1281.7 & 1413.5 & 1469.2 & 288.1 & 23660.1 & 178.5 & 1215.4 \\
\hline
\end{tabular} 
}
\caption{ Total runtime (s) of each data set and measure. Both IW-SSIM and GMSD only evaluate grayscale images. The measures are implemented in Matlab, $^P$ denotes the implementation to be Python based. The number of images of each data set is provided in brackets next to the data set name. The fastest runtime is highlighted in bold.}
\label{tab:runtime}
\end{table*}

\begin{table*}[h]
\centering
\begin{tabular}{c | c c c c | c c c  c}
\hline
  & \multicolumn{4}{c}{Natural Images} & \multicolumn{4}{c}{Medical Images}\\
\hline
                & LIVE$^*$        & LIVE     &  LIVE$^*_{M}$      & LIVE$_{M}$ & PA        & CXR & CT & CT$_{full}$\\
\hline

PSNR & 863 & 890 & 887 & 931 & 857 & 1016 & 860 & 860 \\
SSIM & 884 & 949 & 930 & 1161 & 868 & 1295 & 874 & 880 \\
HaarPSI$^P_{MED}$& \textbf{461} & \textbf{496} & \textbf{500} & \textbf{599} & \textbf{442} & \textbf{854} & \textbf{452} & \textbf{454} \\
HaarPSI$_{MED}$& 879 & 884 & 878 & 920 & 881 & 962 & 883 & 885 \\
MS-SSIM & 873 & 922 & 894 & 1069 & 868 & 1032 & 873 & 869 \\
IW-SSIM & 964 & - & 979 & - & 961 & 1042 & 962 & 964 \\
GMSD & 965 & - & 983 & - & 963 & 1082 & 963 & 967 \\
FSIM & 968 & 988 & 985 & 1040 & 961 & 1080 & 963 & 967 \\
MDSI & 996 & 989 & 1053 & 1043 & 968 & 1400 & 969 & 974 \\
LPIPS$^P_{Alex}$ & 899  & 898  & 899  & 898 & 898  & 1698 & 899  & 898  \\
DISTS & 849 & 855 & 851 & 856 & 850 & 875 & 851 & 852 \\
\hline
\end{tabular} 
\caption{
Median RAM (MB) usage over 100 images. Both IW-SSIM and GMSD only evaluate grayscale images. The measures are implemented in Matlab, $^P$ denotes the implementation to be Python based. The least RAM usage is highlighted in bold.}
\label{tab:memory}
\end{table*}